\documentclass[lettersize,journal]{IEEEtran}
\usepackage[utf8]{inputenc}
\usepackage[T1]{fontenc}
\usepackage{textcomp}
\usepackage{ragged2e}

\usepackage{graphicx}
\usepackage{amsmath}
\usepackage{amssymb}
\usepackage{booktabs}
\usepackage{tabularx}
\usepackage{array}  
\usepackage{makecell}   
\usepackage{longtable}  
\usepackage{caption}
\usepackage{enumitem}   
\usepackage{float}      
\usepackage{placeins}   
\usepackage{hyperref}
\usepackage{xr}

\begin{document}

\title{Modeling Political Discourse with Sentence-BERT and BERTopic}

\author{
\IEEEauthorblockN{Margarida Mendonça and Álvaro Figueira}\\[0.5em]
\IEEEauthorblockA{Department of Computer Science, Faculty of Sciences, University of Porto, Portugal\\
\texttt{arfiguei@fc.up.pt}}
}

\maketitle

\begin{abstract}
Social media has reshaped political discourse, offering politicians a platform for direct engagement while reinforcing polarization and ideological divides. This study introduces a novel topic evolution framework that integrates BERTopic-based topic modeling with Moral Foundations Theory (MFT) to analyze the longevity and moral dimensions of political topics in Twitter activity during the 117th U.S. Congress. We propose a methodology for tracking dynamic topic shifts over time and measuring their association with moral values and quantifying topic persistence. Our findings reveal that while overarching themes remain stable, granular topics tend to dissolve rapidly, limiting their long-term influence. Moreover, moral foundations play a critical role in topic longevity, with Care and Loyalty dominating durable topics, while partisan differences manifest in distinct moral framing strategies. This work contributes to the field of social network analysis and computational political discourse by offering a scalable, interpretable approach to understanding moral-driven topic evolution on social media.


\end{abstract}


\section{Introduction}
Social media has revolutionized society, notably in politics, where it drives civic engagement, increases voter turnout, and mobilizes grassroots movements (\cite{Satterfield2020}). 
It also enables politicians to campaign directly, informally, and at a lower cost, fostering dialogue between elected officials and constituents (\cite{tedx_talks_how_2018}).
Among social media platforms, Twitter (X as of July 2023) is the most adopted by politicians (\cite{cycles_topic_nodate}). 
Its 280-character limit promotes brevity, while its real-time nature enables rapid information sharing. Hashtags and trending topics facilitate tracking discussions, and consistent engagement enhances political visibility and branding. As a result, Twitter has become a key driver of political success (\cite{reveilhac_impact_2023}) and an important source of user-generated data.

A notable example is Donald Trump’s 2016 presidential campaign, which leveraged data mining and analytics, including A/B testing, voter behavior forecasting, and geo-targeting, to engage specific demographics (\cite{noauthor_one_nodate}). 
NLP provided real-time insights into public sentiment, shaping campaign messaging. Techniques like Topic Mining identified emerging themes, while Sentiment Analysis gauged public reactions.

Twitter’s downsides, particularly its role in spreading disinformation and polarization, are well documented (\cite{CenterHumaneTech2021}). While Sentiment Analysis has been used to study these effects (\cite{ozturk_sentiment_2018,bor_quantifying_2023}), Moral Foundations Theory (MFT) remains a lesser-known framework\footnote{https://www.mft-nlp.com/papers.html}. 
It suggests that moral reasoning is shaped by innate, cross-cultural foundations influencing political ideology (\cite{graham_liberals_2009}). 
On Twitter, users often form echo chambers reinforcing their moral perspectives, deepening political polarization (\cite{wronski_intergroup_2016}). 

This work was motivated by the interest in obtaining a global perspective of Congress topics. Topic Mining is often conducted either at the individual level, focusing on a specific person (\cite{hajjej_trump_2020}), 
or at a broader level, where a particular theme is analyzed (\cite{zhou_guided_2023}). 
Positioned at an intermediate level, this research examines Congress as a whole, specifically the 117$^{th}$ session. By providing a comprehensive frame of reference on the key concerns shaping U.S. politics at that time, it lays the groundwork for future comparative analyses. In particular, it establishes a foundation for a direct comparison with the 119$^{th}$ Congress, enabling an assessment of how political discourse has evolved, particularly over the Trump presidential mandates. The forthcoming analysis of the 119$^{th}$ Congress will be presented in a separate publication, allowing for a detailed and structured comparison.  
Building upon previous work, the present analysis explores topic extraction and general trends in the 117$^{th}$ Congress. The interested reader can consult additional details in \cite{mendonca2024bertopic}. The research questions this work aims to answer are as follows:

\noindent \textbf{RQ1:} What topics define the 117$^{th}$ US Congress? 
\hangindent=2.7em \hangafter=1 How do these topics behave throughout the duration of Congress? Are there differences in the duration of these topics?

\noindent \textbf{RQ2:} Is there a relation between topic duration and the Moral Foundations they display? 
\hangindent=2.7em \hangafter=1 How can topic duration be interpreted under Morality Foundations Theory? Is there a significant difference in Foundations across party lines? How can this knowledge help foster political success on Twitter?


\section{Background}

Given the context of this work, this Chapter aims to ensure the reader has the essential knowledge of U.S. politics. We provide some background information on its design, specifically on Congress's structure and functioning. A brief overview of U.S. politicians' activity on Twitter is also delivered. 

\subsection{A Brief Contextualization of U.S. Politics}

The United States political system encompasses three branches: Legislative, comprised of Congress, and responsible for law-making; Executive, which implements laws and policies and is led by the U.S. President; and Judicial, composed of the Supreme Court and smaller federal courts, which ensure laws are abided.
The U.S. political landscape contains over 400 different parties, but it is dominated by only two nationally recognized ones. The Democratic Party platform is social liberalism 
and it believes "\textit{that the economy should work for everyone, health care is a right, our diversity is our strength, and democracy is worth defending} \cite{DemocraticParty_where_nodate}. 
The Republican Party, also known as GOP ('Grand Old Party), runs on a conservative ideology and stands "\textit{for freedom, prosperity, and opportunity [...]. The principles of the Republican Party recognize God-given liberties while promoting opportunity for every American}" \cite{GOP}. 

\textbf{\textit{We The People}}.
The U.S. Constitution has been in operation since 1789 and is the longest-surviving written charter of government in existence \cite{constitution}. 
Article I of the Constitution creates a Bicameral Congress consisting of a Senate and a House of Representatives. This solution came as a result of the Great Compromise. At the time of writing the Constitution, Framers - those involved in drafting it - had conflicting positions on defining State representation. Framers from large States defended representation proportional to population, while those from small States argued for equal representation. As a compromise, Congress was separated into two Chambers \cite{Benzine2015_bicameral}. 

The House of Representatives consists of 435 members, elected every two years, with seats proportional to each State's population. It can impeach officials, decide presidential elections if no candidate wins the electoral majority, and initiate tax bills. Key roles include the Speaker of the House, its highest-ranking member, the Majority Leader, and the Minority Leader.

The Senate has 100 members, two per State, with one-third elected every two years for six-year terms. It conducts impeachment trials, ratifies treaties, and confirms appointments. The U.S. Vice-President serves as Senate President, while the Majority and Minority Leaders oversee legislative matters \cite{Benzine2015}.

The 117$^{th}$ Congress convened on January 3, 2021, following the 2020 elections. Democrats retained House control, with Speaker Nancy Pelosi, Majority Leader Steny Hoyer, and Minority Leader Kevin McCarthy. After Joe Biden's inauguration, Vice-President Kamala Harris became President of the Senate, giving Democrats control with Majority Leader Chuck Schumer and Minority Leader Mitch McConnell. Major events of the 117$^{th}$ Congress include the January 6$^{th}$ Capitol attack, Trump’s impeachment, U.S. sanctions on Russia over Ukraine, Ketanji Brown Jackson’s Supreme Court nomination, and the overturn of Roe v. Wade. Congress also passed major bills, including the Inflation Reduction Act, Infrastructure and Jobs Act, CHIPS and Science Act, Honoring our PACT Act, and the Respect for Marriage Act \cite{Binder2022}.

\subsection{American Politics and Twitter/X}
As of 2022, Congress members total of 515 Twitter accounts, but not all accounts are used the same. Firstly, some members have personal accounts and professional accounts. The former tends to be more informal, with the member managing it directly, while a team runs the latter. Secondly, Democrats tend to tweet more and have more followers, but engagement is evenly split among parties \cite{Lee2019}.  
There are some members who have taken to Twitter better than others \cite{Mills2021}.

\section{Materials and Methods}
The present section aims to briefly introduce the concepts of Topic Extraction, Moral Foundations Theory and Topic Longevity. It also presents the steps performed for data selection, extraction and preprocessing. 

\subsection{Topic Extraction with BERTopic}
A well-known fact in the data science community is that the rise of the Internet and social media has led to an explosion of data. Extracting valuable insights from this requires proper manipulation and analysis. Textual data accounts for approximately 75\% of all web content~\cite{noauthor_what_nodate}, posing unique processing challenges. It is unstructured, lacks a schema, and is often ambiguous—using synonyms, slang, and abbreviations. Contextual understanding is crucial, while spelling errors and poor grammar further complicate analysis. Social media exacerbates these challenges due to its informal, short-text nature.

Natural Language Processing (NLP) enables computers to process and generate human language. Among its tasks is Topic Mining, which uncovers and categorizes key themes in a document without prior knowledge. Advances in Machine Learning (ML) have significantly improved topic modeling, with Transformers emerging as a powerful approach~\cite{vaswani_attention_2023}. A key feature of Transformers is the self-attention mechanism, which weighs word relevance while considering their relationships. This enables them to capture both local and global dependencies, making them particularly effective for tasks like translation, text generation, and language understanding. In 2020, BERTopic was introduced as \textit{“a topic model that leverages clustering techniques and a class-based variation of TF-IDF to generate coherent topic representations”}~\cite{grootendorst2022bertopic}. It addresses prior models' limitations, particularly their inability to capture semantic relationships between words.

\begin{figure}[!t]
    \centering
    \includegraphics[width=0.8\columnwidth]{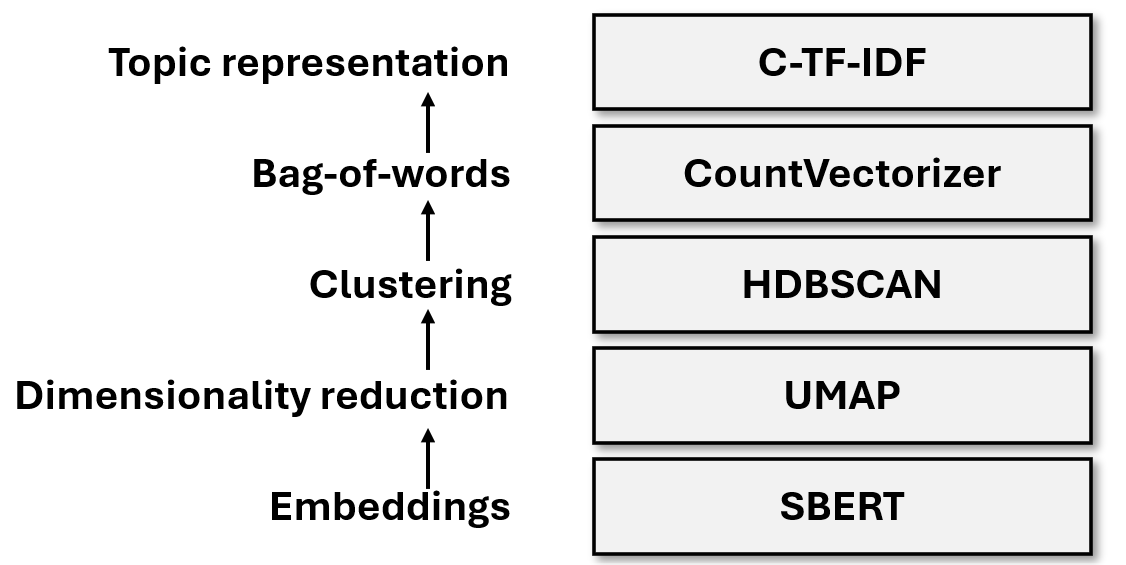}
    \caption{BERTopic algorithm}
    \label{fig:bert-algo}
\end{figure}

BERTopic mines topics in five stages (Fig. \ref{fig:bert-algo}, from \cite{grootendorst_algorithm_nodate}), where we present our choices for each stage. 
Although the default algorithms have been selected for the reasons presented below, BERTopic is highly modular, and users can customize it at each step. It also allows the fine-tuning of each default algorithm through its corresponding hyperparameters. The algorithm begins by converting documents into embedding representations. The default algorithm for this is Sentence-BERT, which achieves state-of-the-art performance on such tasks (\cite{reimers_sentence-bert_2019}). The dimensionality of these embeddings tends to be quite high, with some models achieving ten thousand dimensions (\cite{james_briggs_bertopic_2022}). UMAP is used to reduce this to 2D or 3D since it preserves local and global features of high-dimensional data better than alternatives such as PCA or t-SNE (\cite{statquest_with_josh_starmer_umap_2022}). With data in a more feasible vector space, it can now be clustered. HDBSCAN allows noise to be considered outliers, does not assume a centroid-based cluster, and therefore does not assume a cluster shape - an advantage to other Topic Modeling techniques (\cite{statquest_with_josh_starmer_clustering_2022}). The next step is to perform c-TF-IDF. This is a variation of the classical TF-IDF: firstly, generate a bag-of-words at the cluster level, concatenating all documents in the same class. Next, apply TF-IDF to each cluster bag-of-words, resulting in a measure for each cluster, instead of a corpus-wide one. 

BERTopic’s default embedding model is SBERT, a transformer-based approach that outperforms traditional methods. SBERT provides various pre-trained models, including multilingual and multi-modal options. For our dataset, "all-MiniLM-L6-v2" was used.  
By default, UMAP performs dimensionality reduction in BERTopic, but PCA can be chosen for simpler datasets, often offering better speed. Similarly, HDBSCAN is the default clustering algorithm, but we also compare it with k-Means, which does not allow outliers. HDBSCAN, however, is more robust for high-dimensional data like ours.  

The Vectorizer stage employs CountVectorizer, with customizable parameters tailored to the dataset, such as n-grams to define token length in topics. While BERTopic’s c-TF-IDF inherently removes low-frequency words, early removal optimizes the topic-term matrix and speeds up processing.  

The c-TF-IDF for a term x in class c is given by:  
\begin{equation}
    W_{x,c} = \lVert tf_{x,c}\rVert \cdot \log(1+\frac{A}{f_x})
\end{equation}
where $\textbf{tf}_{x,c}$ is the frequency of word $\textbf{x}$ in class $\textbf{c}$, $\textbf{f}_c$ is the frequency across all classes, and $\textbf{A}$ is the average word count per class. Two key parameters fine-tune c-TF-IDF:  
- \textit{bm25\_weighting} (boolean) adjusts the weighting scheme, with BM25 defined as:  
  \begin{equation}
  \log(1+\frac{A - f_x + 0.5}{f_x + 0.5})
  \end{equation}
  This is particularly effective for smaller datasets containing stopwords.  
- \textit{reduce\_frequent\_words} (boolean) applies a square root to term frequency after normalization:  
  \begin{equation}
  \sqrt{\lVert tf_{x,c}\rVert}
  \end{equation}

Combining both parameters results in:
\begin{equation}
    \begin{aligned}
        W_{x,c} &= \sqrt{\lVert tf_{x,c}\rVert} \cdot \log(1+\frac{A - f_x + 0.5}{f_x + 0.5})
    \end{aligned}
\end{equation}

Since topic modeling evaluation is inherently challenging, we rely on perplexity measures, topic coherence, and topic diversity to assess BERTopic performance. The following table outlines our optimized BERTopic settings:

\begin{table}[!t]
    \caption{Optimized BERTopic Parameters}
    \label{tab:bertopic_selected_options}
    \centering
    \begin{tabular}{
        >{\raggedright\arraybackslash}p{3.2cm} 
        >{\raggedright\arraybackslash}p{5.8cm}}
        \toprule
        \textbf{Stage} & \textbf{Selected Parameters} \\
        \midrule
        Embeddings & Word2Vec \\
        Dimensionality Reduction & Base Dimensionality Model \\
        Clustering & HDBSCAN \\
        Vectorizer & \texttt{ngram\_range} = (1,1), \texttt{min\_df} = 5, \texttt{max\_features} = 170{,}000 \\
        Topic Representation & \texttt{bm25\_weighting} = False, \texttt{reduce\_frequent\_words} = True \\
        \bottomrule
    \end{tabular}
\end{table}

\subsection{Longevity}
To evaluate how topics evolve, it is first necessary to develop a measure of linkage between topics of consecutive months. Topics are characterized by their representations, i.e., words that strongly identify them. As topics change monthly, the representations vary accordingly. Measuring the similarity of consecutive months' topics can be achieved by comparing the corresponding representations. This comparison was done  using cosine similarity. Cosine similarity is the cosine of the angle between two vectors; a cosine similarity of 0 suggests no similarity between documents, while a value of 1 says the documents are the same. Concretely, in the context of document comparison this metric measures the angle of each document's vector. This vector is n-dimensional, where  \textit{n} is the number of words present in the document. 

\begin{figure}[!t]
    \centering
    \includegraphics[width=\linewidth]{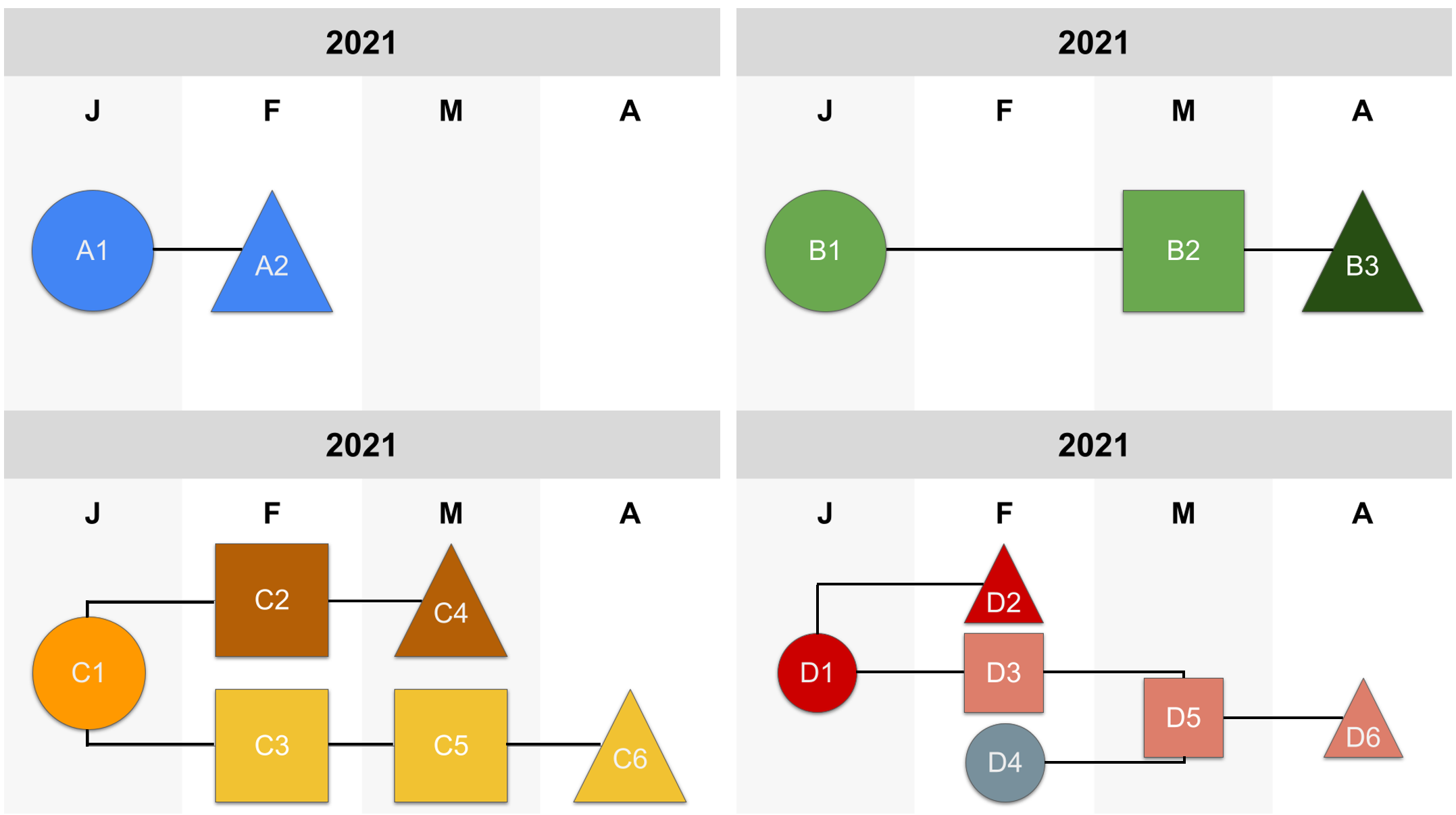}
    \caption{Examples of topic evolution patterns: 
    (a) short-lived topic, 
    (b) topic with intermittent activity, 
    (c) topic splitting into two, and 
    (d) long-lived merged topic.}
    \label{fig:case1234}
\end{figure}

Topic evolution is determined by calculating the cosine similarity for all consecutive months and reducing the topic selection to those with non-negative similarity. 
In terms of visualization (Fig. \ref{fig:case1234}), each vertical bar represents a unique month, where multiple topics can appear.  Topics are represented by one of three symbols: circles indicate topic emergence, squares are for topic stagnation, and triangles are for topic disappearance. Similar colors of consecutive topics represent a similarity above 0.5; topic splits or mergers are highlighted by changes in colors. Additionally, the different stages that compose the evolution of the same topic are attributed a unique identifier composed of a capital letter and a number. 

Fig.\ref{fig:case1234} illustrates four fictional cases from 2021. Topic A (blue) is the shortest, lasting two stages with a longevity of 2. It emerges in January (A1, circle) and disappears in February (A2, triangle). Topic B (green) spans three stages but has a longevity of 4, from January to April. B2 represents stagnation, while B3’s color change reflects a diminishing coefficient. Topic C (yellow) exemplifies topic splitting, where C1 evolves into two branches. Despite differing longevities, the longest branch determines Topic C’s overall longevity of 4. Finally, Topic D (red) demonstrates a topic merger. Stages D3 and D4 combine, extending Topic D's longevity to 4. Since D4 is unrelated to previous stages, its color differs.

\subsection{Moral Foundations Theory}
The Moral Foundations Theory (MFT) is a framework designed by Jonathan Haidt and Jesse Graham  (\cite{graham_liberals_2009}). Its primary goal is to explore the presence of common moral themes across cultures. MFT suggests the existence of inherent psychological systems that form the basis of intuitive ethical judgments. Cultures then build upon these systems to develop virtues and societal structures, resulting in the diverse moral beliefs observed globally. This paradigm of human morality is descriptive, meaning it refrains from making normative claims about the moral goodness of these systems. The original MFT framework identified five foundational pillars: 

\begin{enumerate}
    \item \textbf{Care/Harm}: This foundation is related to humans' evolution as mammals, with attachment systems and an ability to feel the pain of others. It underlies the virtues of kindness, gentleness, and nurture.
    
    \item \textbf{Fairness/Proportionality}: Related to the evolutionary process of reciprocal altruism, which is a concept that suggests individuals in a community can benefit from cooperating with one another. In the context of morality, it suggests that humans have evolved to value principles of justice and rights because, historically, cooperating and treating others fairly led to mutual benefits and enhanced their survival and well-being.

    \item \textbf{Loyalty/Disloyalty}: This foundation is related to Human history as tribal creatures that created coalitions. It underlies the principles of patriotism and self-sacrifice for the greater good. 
    
    \item \textbf{Authority/Subversion}: This foundation was shaped by the history of hierarchical social interactions. It underlies virtues of leadership, including deference to prestigious authority figures and respect for traditions.

    \item \textbf{Purity/Degradation}: It is influenced by the psychological concepts of disgust and contamination and it is related to the human inclination to aspire to a less primal, more dignified way of living, a notion often present in religious narratives.  This foundation supports the belief that the body is sacred and can be defiled by immoral actions and impurities.
\end{enumerate}

Moral Foundations Theory is broadly adopted, in part due to the development of the Moral Foundations Dictionary (MFD) (\cite{graham_liberals_2009}). 
The MFD defines a taxonomy of values together with a term dictionary, making it an important resource for NLP applications. Since Moral Foundations are focused on intrinsic psychological systems, they can be assigned to groups by their written documents, specifically, party members' tweets. Hence, our hypothesis sequence is: $ \text{Party} \rightarrow \text{Moral Foundations} \rightarrow \text{Longevity} $.

\noindent \textbf{H1:} Longevity of a tweet is related to an author's party. 
\hangindent=2.3em \hangafter=1 Democrats' tweet longevity is statistically significantly different from that of Republicans. For simplicity, we discard Independents. 

\noindent \textbf{H2:} Affiliation to different parties translates into different morals. 
\hangindent=2.3em \hangafter=1 This has been suggested in the literature, and we expect the data to confirm this \cite{koleva_tracing_2012}.

\noindent \textbf{H3:} Longevity is driven, at least in part, by the Moral Foundations detected in tweets. 
\hangindent=2.3em \hangafter=1 Assuming H1 and H2 hold, there should be statistically significant differences in longevity based on morals.

\subsection{Data Selection and Extraction}

This work analyzes Twitter activity from the 117$^{th}$ U.S. Congress, covering 102 Senators and 413 Representatives, totaling 515 accounts. To reduce this number, we first assessed their activity. \hyperlink{https://twitter.com/tweetcongress}{Tweet Congress}, a grassroots project tracking congressional Twitter usage, was used for data selection. Though no longer available, it provided a list of the top 10 most active users.

Key congressional figures—majority and minority leaders—were also included. Party representation was balanced, and both personal and professional accounts were considered. Table SM.1 in the Supplementary Material lists the final selection: 27 members with 40 accounts (13 Democrats, 13 Republicans, one Independent). Eleven are Representatives, 14 are Senators, and two are the U.S. President and Vice-President. Tweets from these accounts were extracted via the Twitter API (v1.1), covering the entire 117$^{th}$ Congress, totaling 27,782 tweets.

\subsection{Data Preprocessing}
Several preprocessing steps were applied using various Python packages that support NLP preprocessing, including NLTK, SpaCy, and Stanford NLP. Given the task's simplicity, any could be used, but NLTK was chosen for its ease of use, rich resources, and good maintenance.

Tokenization and cleaning involved replacing web characters (e.g., \textit{$<$br$>$}, \textit{\%quot;}, \textit{\&\#39;}, \textit{\&amp;}) with their intended meaning, removing hashtags, hyperlinks, and mentions while creating new features, eliminating redundant "RT" tags, stripping punctuation, and discarding empty tweets.  
Beyond these new features, the dataset included \textit{color} (author’s party: Democrat (blue), Republican (red), Independent (white)) and \textit{account type} (professional or personal). After cleaning, stopwords from the NLTK corpus were removed, and lemmatization applied. Lemmatization was preferred over stemming, as short-text data has limited context, and further reduction could compromise meaning. Tweets emptied by stopword removal were discarded.

\section{Results}
\begin{figure*}[!t]
    \centering
    \includegraphics[width=\textwidth]{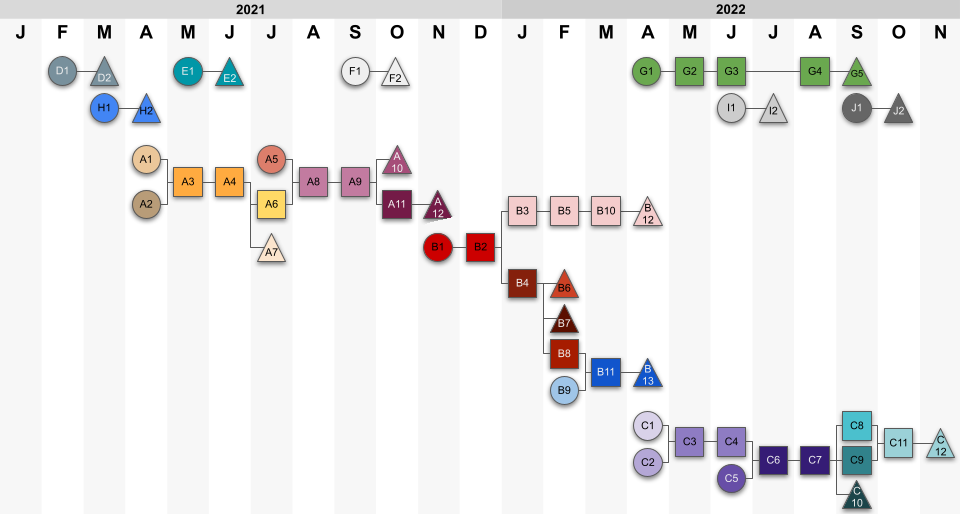}
    \caption{Timeline of topic evolution across the 24-month period. Each horizontal bar represents a topic, spanning the months during which the topic was active. Long bars indicate persistent topics (broad issues recurring over time), whereas short bars represent ephemeral topics tied to transient events.}
    \label{fig:topic-timeline}
\end{figure*}

Our initial analysis provided a global view of Congress topics, but social media evolves rapidly. 
To capture this evolution, we reduced the interval for topic extraction. 
A monthly scale was chosen to balance detail with sufficient tweet volume. BERTopic runs independently for each bin with optimized stages. Table SM.2 lists the number of non-outlier topics per month and the two most frequent ones. In total, 175 topics were extracted from 13,436 tweets. In January 2021, no topics were found. Generally, the most common topics (Topic 0) align with previous findings, while the second most frequent introduce new themes. The final dataset contains 54 topics, illustrated in Fig. \ref{fig:topic-timeline}. Consecutive topics with similar colors indicate similarity above 0.5, while color shifts highlight topic splits or mergers.


The longevity of a topic is the number of months a topic lasted without fully disappearing. Table SM.3 illustrates the respective derived groups. It is possible to identify three particular cases of Longevity in the data:

\noindent \textbf{High Longevity} (more than six months) includes groups A and C. Group A begins in April 2021 discussing \textit{infrastructure}, \textit{democrats}, and \textit{republicans}. In July, the topics split into \textit{marijuana}, \textit{family payments}, and \textit{democracy, january, facts}. The former two then merge into \textit{tax}. In October, a new split occurs: \textit{americans} and \textit{register, vote, day}. Group C begins in April 2022 with \textit{maternal, women, health} and \textit{oklahoma, unconstitutional, abortion}. These merge into \textit{right, abortion}, which in July also merges with \textit{child, affordable}, resulting in \textit{right, biden, act}. This later splits into \textit{abortion}, \textit{student debt}, and \textit{national, vote, registration}. The group disappears in November with \textit{poll, location, early}.

\noindent \textbf{Medium Longevity} (five to six months) includes groups B and G. Group B begins in November 2021 with \textit{infrastructure, bills, families}. It splits into \textit{workers, organize, corporate}, as well as \textit{drug, prescription} and \textit{economy, billionaires}. Topic emergence occurs simultaneously with the split: \textit{russia, ukraine, war}. This stagnates until April 2022. Group G encompasses tweets where Matt Gaetz promotes his podcast Firebrand.

\noindent \textbf{Short Longevity} (less than five months) includes groups D through J. Given their short duration, topics in this group are unlikely to merge or split. Group D mentions \textit{relief checks}, while group E evolves from \textit{israel, palestinians} to \textit{lgbtq, pride, month}. Group F covers Jewish celebrations and later the release of Adam Schiff's book \textit{Midnight in Washington}. Group H discusses \textit{vote, georgia, capital, evans}, group I focuses on \textit{prime, minister, abe}, and group J on \textit{israel, nuclear, lebanon}.

To analyze the moral foundations present in the data, the package \textit{moralstrength} was used, as developed \cite{araque_moralstrength_2020}. 
In this paper, the authors expand on the MFD to overcome its limitations, namely, increasing its variety and diversity of lemmas, as well as developing a scale of strength of Foundations, instead of only a binary metric.


\noindent \textbf{H1: Longevity of a tweet is related to an author's party.} 
To explore whether Longevity and Color are related, we first examine their distributions. Both groups exhibit similar profiles, with Democrats showing a slightly higher frequency around Short Longevity. Given this similarity, we proceed directly to selecting the appropriate statistical test of independence.

Longevity is a categorical, ordinal variable, and Color is categorical. Both populations, Democrats and Republicans, are assumed to be independent. Given these conditions, a $\chi^2$ test is used. In a $\chi^2$ test, the null hypothesis states that the two variables are independent. To reject it, the p-value must be below the significance level, $\alpha$. Running the test, we obtain a p-value of 1, the highest possible. Thus, H1 cannot be verified.

\begin{figure*}[!t]
    \centering
    \includegraphics[width=\textwidth]{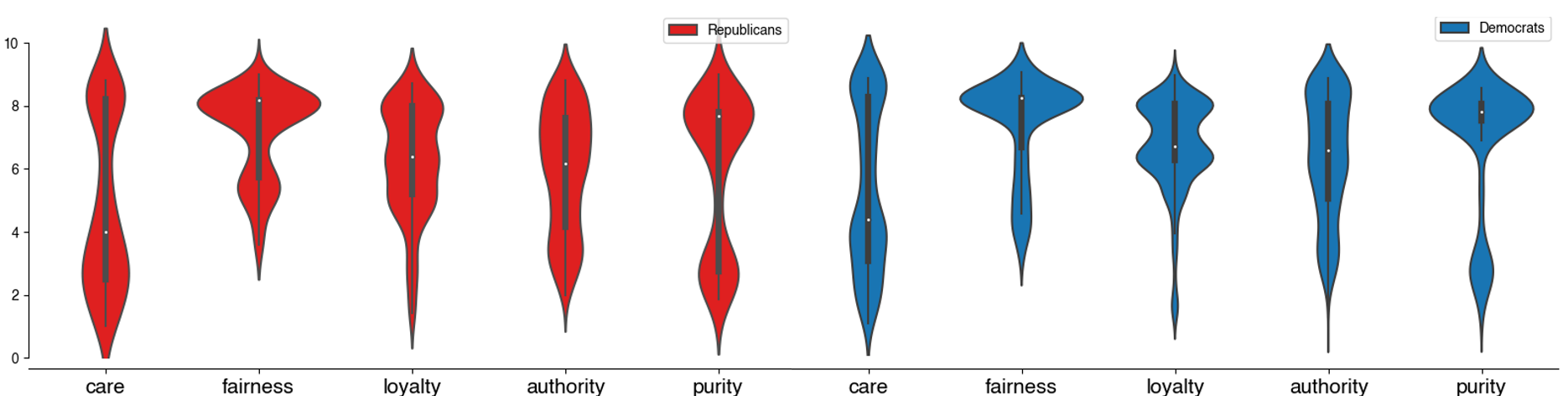}
    \caption{Distribution of moral foundation scores for Democratic and Republican legislators. Violin plots show distinct partisan emphasis: Democrats exhibit higher scores on Care and Fairness, while Republicans show elevated scores on Loyalty, Authority, and Purity.}
    \label{fig:morality-violin}
\end{figure*}

\noindent \textbf{H2: Affiliation to different parties translates into different Moral Foundations.}  
A similar strategy is followed to investigate H2. Firstly, Fig. \ref{fig:morality-violin} illustrates the distributions of MF scores according to color. It suggests Republican tweets tend to score lower in Care, Authority, and Purity. Fairness scores are similar. Loyalty has a slightly larger frequency of lower scores in Republicans.

Moral Foundations are continuous variables, while color is categorical, with both populations assumed independent. To assess correlation, we use the Mann-Whitney U test. Table \ref{tab:mannwithney-mf} presents the p-values for each MF between parties. At a 95\% significance level, the null hypothesis of no relationship is rejected; at 99\%, this holds for all except Fairness.

\begin{table}[!t]
    \caption{Mann–Whitney U test statistics comparing average moral foundation scores between Democratic and Republican legislators}
    \label{tab:mannwithney-mf}
    \centering
    \begin{tabular}{lccc}
        \toprule
        \textbf{Moral Foundation} & \textbf{p-value} & \textbf{Dem. Avg.} & \textbf{Rep. Avg.} \\
        \midrule
        Care      & 0.00022  & 5.27 & 4.80 \\
        Fairness  & 0.0137   & 7.36 & 7.17 \\
        Loyalty   & 0.00060  & 6.65 & 6.22 \\
        Authority & 0.0028   & 6.34 & 5.95 \\
        Purity    & 0.00008  & 6.76 & 5.91 \\
        \bottomrule
    \end{tabular}
\end{table}

\begin{figure*}[!t]
    \centering
    \includegraphics[width=\textwidth]{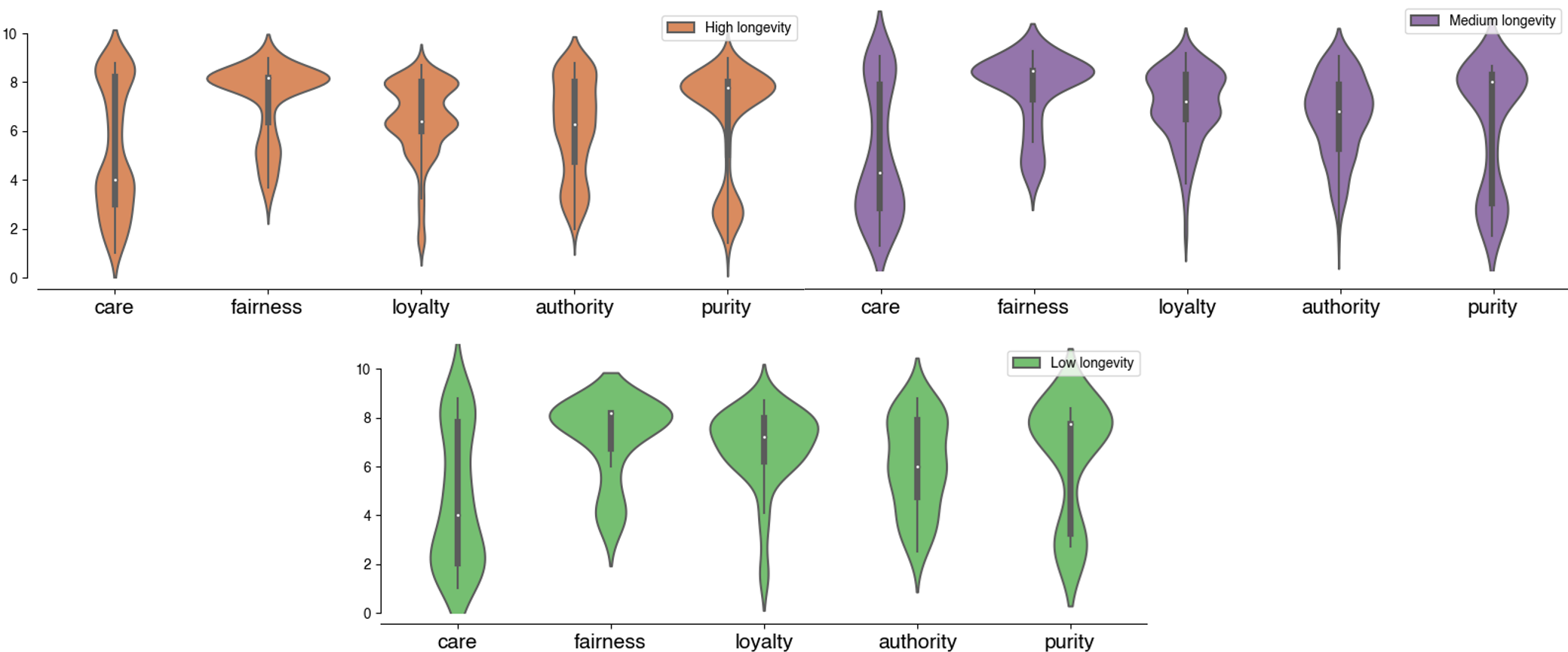}
    \caption{Distribution of moral foundation scores across topics grouped by longevity. Long-lived topics tend to have higher Care and Loyalty scores, while short-lived topics exhibit weaker moral framing.}
    \label{fig:morality-violin-long}
\end{figure*}

\noindent \textbf{H3: Longevity is driven, at least in part, by Moral Foundations detected in tweets.}  
Similarly to before, this hypothesis compares a categorical variable (Longevity) with a continuous one—Moral Foundations. Fig. \ref{fig:morality-violin-long} illustrates the distribution of MF scores for the three levels of Longevity. Medium and Short Longevity share similar distributions for all foundations except Loyalty. Since Longevity has more than two levels, the Mann-Whitney U test was run for each pair of MF across Longevity levels.  

Table \ref{tab:mannwithney-mf-long-all} presents the p-values of the Mann-Whitney U test for each MF, combining Low and Medium Longevity into one (additional combinations appear in Supplementary Material, Table SM.4). The lowest p-values were obtained for Care and Loyalty: at a 99\% significance level, the null hypothesis of no relationship is rejected for these but not for the remaining foundations.

\begin{table}[!t]
    \caption{Mann–Whitney U test results comparing average moral foundation scores across topics with different longevity levels}
    \label{tab:mannwithney-mf-long-all}
    \centering
    \begin{tabular}{lccc}
        \toprule
        \textbf{Moral Foundation} & \textbf{p-value} & \textbf{Long-lived Avg.} & \textbf{Short-lived Avg.} \\
        \midrule
        Care      & 0.0009  & 5.21 & 4.69 \\
        Fairness  & 0.7817  & 7.27 & 7.37 \\
        Loyalty   & 0.0072  & 6.48 & 6.76 \\
        Authority & 0.7023  & 6.15 & 6.27 \\
        Purity    & 0.2059  & 6.47 & 6.03 \\
        \bottomrule
    \end{tabular}
\end{table}

\section{Discussion}
The monthly binning allowed for an increase of 230\% in tweets assigned to a topic. This implies that while the most common topics found for each month tend to align with the overarching ones, there is more granularity found for the subsequent topics. In essence, the most common topics highlight the deeper, structural topics Congress faced (Abortion, StudentDebt), while the remaining relate to singular events such as the Capitol Car Attack or the Assassination of Former Prime Minister Abe Shinzo. While a weekly binning might arguably bring even more detail, the data was not sufficient for topic extraction. Broader topics also evolve through time, but not simultaneously. Data suggests that High Longevity topics occur one at a time and that the end of one comes with the beginning of another. 

Regarding the investigation of Moral Foundations and Topic Longevity, the statistical analysis indicates tweeters of different parties tend to score differently in Moral Foundations. As previously mentioned, this framework is merely descriptive, not providing any sense of superiority of one system over the other. On average, Republicans score lower in Care than Democrats. Care's characteristic emotion is compassion (\cite{schuman_moral_2018}) 
and an example of tweets with distinct levels are:
\begin{quote}
\textit{“The Senate just passed a bipartisan infrastructure bill. That’s a big deal. But the next step must be an investment in our people, too. We need to address climate, health care, housing, education and childcare. Investing in our people, our country, and our planet. Let’s do it.”} \\
\hfill — @RepAndySchiff, 10-08-2021 (Democrat) \quad \textbf{Care score: 8.8}
\end{quote}

\vspace{1ex}

\begin{quote}
\textit{“More than 3 million people have entered our country illegally on Biden’s watch. Some are on the terrorist watch list and others are trafficking fentanyl, which is now the greatest killer of Americans aged 18–45.”} \\
\hfill — @GOPLeader, 03-10-2022 (Republican) \quad \textbf{Care score: 1.0}
\end{quote}

Loyalty scores of Democrats are higher on Loyalty than Republicans. This Foundation is related to the in-group/out-group perspective. This result differs from those found by previous literature, suggesting a change of attitudes from Democrats:
\begin{quote}
\textit{“Under Trump and DeVos, the victims of predatory for-profit colleges were abandoned. Now @POTUS and @SecCardona are helping student loan borrowers get debt relief faster and simpler—and forcing colleges to cover the cost of fraud. Special interests lost.”} \\
\hfill — @SenWarren, 01-11-2022 (Democrat) \quad \textbf{Loyalty score: 1.6}
\end{quote}

\vspace{1ex}

\begin{quote}
\textit{“The illegitimate January 6 Committee's vote to subpoena President Trump is a political hatchet job read by a political hatchet committee. This committee is illegitimately formed, in violation of House rules, and is organized to search and destroy perceived political enemies.”} \\
\hfill — @RepAndyBiggsAZ, 13-10-2022 (Republican) \quad \textbf{Loyalty score: 8.71}
\end{quote}

Republicans score lower in Authority than Democrats. This Foundation applies to traditions, institutions, and values; typically Social Conservatives tend to score higher than Social Liberals. However, this lower score can be interpreted as against the Authority of the ruling party:

\begin{quote}
\textit{“No amnesty for COVID bureaucrats. No amnesty for illegal aliens.”} \\
\hfill — @RepAndyBiggsAZ, 05-11-2022 (Republican) \quad \textbf{Authority score: 3.2}
\end{quote}

\vspace{1ex}

\begin{quote}
\textit{“The former president is still trying to stonewall subpoenas. But this time, we have a Justice Department devoted to the rule of law. This time, lawbreaking witnesses must weigh the prospect of criminal prosecution. Americans deserve answers. We will make sure they get them.”} \\
\hfill — @RepAdamSchiff, 07-20-2021 (Democrat) \quad \textbf{Authority score: 8.67}
\end{quote}

Purity is correlated with a sense of disgust, and Social Conservatives particularly rely on this Foundation when discussing the sanctity of life (in the abortion debate), the sanctity of marriage (in the gay rights debate), and the sanctity of self (in the contraception debate). Republicans scored lower than Democrats:

\begin{quote}
\textit{“The fight to protect voting rights is a long march and an uphill battle. But nothing is more important than securing the most sacred right in our democracy. We will not give up.”} \\
\hfill — @chuckschumer, 20-01-2022 (Democrat) \quad \textbf{Purity score: 8.13}
\end{quote}

\vspace{1ex}

\begin{quote}
\textit{“These sick individuals should have never been in our country in the first place. The Biden Administration should be enforcing the immigration laws on our books instead of giving a free pass to those unlawfully entering our country.”} \\
\hfill — @RepAndyBiggsAZ, 22-11-2022 (Republican) \quad \textbf{Purity score: 2.71}
\end{quote}

The statistical analysis also confirmed that there is a relationship between Longevity and Moral Foundations Care and Loyalty. High Longevity tweets tend to score higher in Care and lower in Loyalty. 

\begin{figure}[!t]
    \centering
    \includegraphics[width=0.9\linewidth]{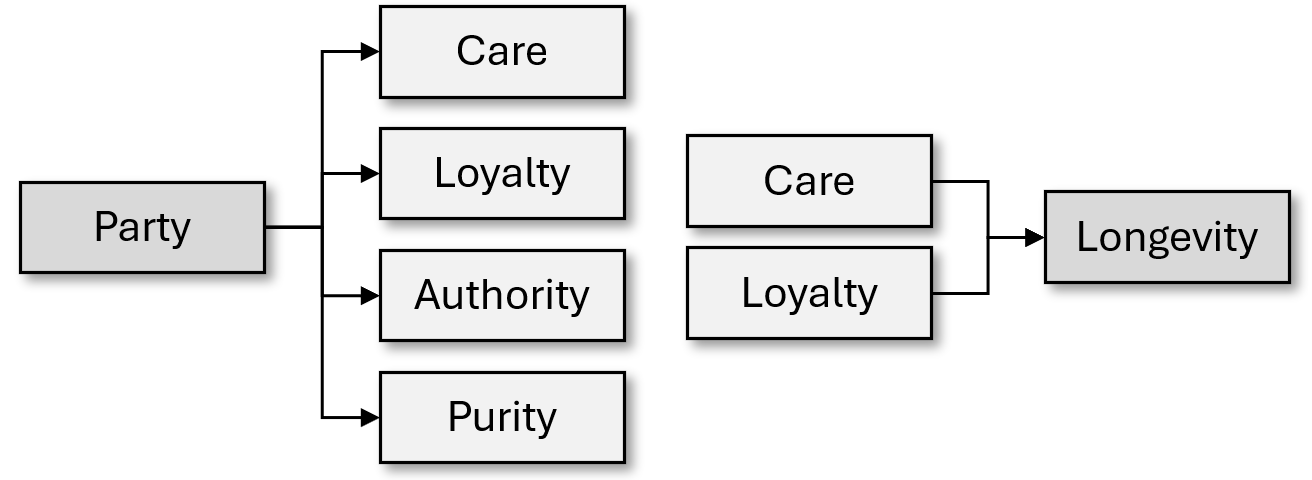}
    \caption{Summary of hypothesis validation results. Hypothesis H1 (moral framing is associated with topic longevity) and H2 (moral framing differs by party affiliation) were both supported by the data.}
    \label{fig:verified-hypothesis}
\end{figure}

Finally, while only two of the hypotheses were confirmed, and the middle connection was not found, tweets that appeal to Care and Loyalty seem to belong to longer-lasting topics. Fig. \ref{fig:verified-hypothesis} illustrates the final conclusions.

\section{Conclusions}
It is now possible to answer the research questions raised at the outset of this work.
Firstly, we show that some topics evolve throughout time, and others disappear. We see that while overarching topics can be found with a global analysis, individual topics can only be identified with a more granular perspective. These, however, tend to dissolve, not necessarily contributing to the conversations of more durable topics. 

The Moral Foundations Theory suggests these behaviors are related to an increased use of Care and Loyalty Foundations. It also hints that differences in political ideology are revealed in distinct uses of Care, Loyalty, Authority, and Purity Foundations.

The main objective moving forward is to use this work as a foundation for a comparative analysis with the 118th and 119th U.S. Congresses, enabling an assessment of how political discourse evolves over time. Beyond this, additional expansion opportunities include integrating engagement metrics into the extracted data to provide richer insights into topic dynamics. Future work might also complement this quantitative analysis with qualitative evaluations of topic significance and impact.

\vspace{6pt}


\bibliographystyle{IEEEtran}

\clearpage
\onecolumn
\appendix
\section*{Supplementary Material}

This supplementary document provides additional tables referenced in the main manuscript. These include:
\begin{itemize}[leftmargin=1.5em]
    \item The list of selected Twitter accounts from members of the U.S. Congress.
    \item Monthly topic extraction results using BERTopic.
    \item Final topic groupings and their semantic labels.
    \item Moral foundation score comparisons across longevity groups.
\end{itemize}

\captionsetup{font=small}
\setlength{\tabcolsep}{6pt}
\renewcommand{\arraystretch}{1.12}

\section*{A. Twitter Account Selection}

\begin{table}[H]  
\caption{Final selection of Congress Twitter accounts}
\label{tab:selected_twitter}
\centering
\small
\begin{tabularx}{\textwidth}{l>{\ttfamily}X c c}
\toprule
\textbf{Member Name} & \textbf{Twitter Handle(s)} & \textbf{Chamber} & \textbf{Party} \\
\midrule
Adam Schiff & @RepAdamSchiff @AdamSchiff & Senator & D \\
Alexandria Ocasio-Cortez & @AOC @RepAOC & Representative & D \\
Andy Biggs & @RepAndyBiggsAZ & Representative & R \\
Bernie Sanders & @BernieSanders @SenSanders & Senator & I \\
Charles Schumer & @SenSchumer @chuckschumer & Senator & D \\
Cory Booker & @SenBooker @CoryBooker & Senator & D \\
Elizabeth Warren & @ewarren @SenWarren & Senator & D \\
Jim Jordan & @Jim\_Jordan & Representative & R \\
Joaquin Castro & @JoaquinCastrotx & Representative & D \\
Joe Biden & @JoeBiden @POTUS & President & D \\
John Cornyn & @JohnCornyn & Senator & R \\
John Kennedy & @SenJohnKennedy & Senator & R \\
Kamala Harris & @KamalaHarris @VP & Vice-President & D \\
Kevin McCarthy & @GOPLeader & Representative & R \\
Lee Zeldin & @RepLeeZeldin & Representative & R \\
Marco Rubio & @SenRubioPress @marcorubio & Senator & R \\
Marjorie Taylor Greene & @RepMTG & Representative & R \\
Marsha Blackburn & @MarshaBlackburn & Senator & R \\
Matt Gaetz & @RepMattGaetz & Representative & R \\
Mitt Romney & @SenatorRomney @MittRomney & Senator & R \\
Nancy Pelosi & @TeamPelosi @SpeakerPelosi & Representative & D \\
Patty Murray & @PattyMurray & Senator & D \\
Pramila Jayapal & @RepJayapal @PramilaJayapal & Representative & D \\
Rand Paul & @RandPaul & Senator & R \\
Rick Scott & @SenRickScott & Senator & R \\
Steny Hoyer & @LeaderHoyer @StenyHoyer & Representative & D \\
Ted Cruz & @SenTedCruz & Senator & R \\
\bottomrule
\end{tabularx}
\end{table}
\FloatBarrier   

\section*{B. Monthly BERTopic Results}

\small
\setlength{\LTcapwidth}{\textwidth}
\begin{longtable}{c l c >{\RaggedRight\arraybackslash}p{0.66\textwidth}}
\caption{Topics found on monthly intervals}
\label{tab:monthly-topics}\\
\toprule
\textbf{Year} & \textbf{Month} & \textbf{Nbr. Topics} & \textbf{Name of Two Most Frequent Topics}\\
\midrule
\endfirsthead
\toprule
\textbf{Year} & \textbf{Month} & \textbf{Nbr. Topics} & \textbf{Name of Two Most Frequent Topics}\\
\midrule
\endhead
\midrule
\multicolumn{4}{r}{\emph{Continues on next page}}\\
\bottomrule
\endfoot
\bottomrule
\endlastfoot
2021 & Jan & 0 & N/A \\
     & Feb & 2 & 0 trumpcovid relief need; 1 robinhood trade customers business \\
     & Mar & 3 & 0 asian american relief check; 1 voter vote georgia democracy \\
     & Apr & 3 & 0 infrastructure republicans democrats climate; 1 officer capitol evans police \\
     & May & 2 & 0 vote trump democracy republicans; 1 israel terrorist hamas palestinians \\
     & Jun & 2 & 0 vote democrats infrastructure democracy; 1 lgbtq pride month carl \\
     & Jul & 6 & 0 marijuana marijuanajustice federal war; 1 child families payments cut \\
     & Aug & 4 & 0 infrastructure bill tax bipartisan; 1 afghanistan afghan administration biden \\
     & Sep & 2 & 0 need democrats tax must; 1 tovah mar kippur yom \\
     & Oct & 6 & 0 bill americans american democrats; 1 midnight book washington democracy \\
     & Nov & 3 & 0 infrastructure bipartisan bill families; 1 hanukkah happy thanksgiving family \\
     & Dec & 2 & 0 workers work back get; 1 contempt meadows mark january6thcmte \\
\midrule
2022 & Jan & 6 & 0 vote act right democracy; 1 kroger wealth workers greed \\
     & Feb & 13 & 0 putin russia ukraine war; 1 starbucks workers organize corporate \\
     & Mar & 3 & 0 ukraine putin russia work; 1 lord psalms praise matthew \\
     & Apr & 21 & 0 jackson ketanji brown judge; 1 starbucks amazon workers amazonlabor \\
     & May & 3 & 0 right abortion must fight; 1 cuba de en la \\
     & Jun & 17 & 0 abortion roe gun right; 1 student debt cancel potus \\
     & Jul & 2 & 0 right biden get act; 1 abe japan minister prime \\
     & Aug & 3 & 0 inflation act reduction cost; 1 gaetz matt firebrand episode \\
     & Sep & 40 & 0 abortion ban reduction care; 1 strong housegop economy commitment \\
     & Oct & 2 & 0 biden make democrats republicans; 1 israel isaac herzog lebanon breakthrough \\
     & Nov & 30 & 0 poll voice location early; 1 victory np congratulations democratic \\
\end{longtable}
\FloatBarrier

\section*{C. Topic Groups and Labels}

\begin{table}[H]
\caption{Grouped topics and their labels}
\label{tab:grouped-topics}
\centering
\small
\begin{tabularx}{\textwidth}{l X}
\toprule
\textbf{Group} & \textbf{Topics} \\
\midrule
A & 0\_infrastructure\_republicans\_democrats, 0\_vote\_trump\_democracy\_republicans, etc. \\
B & 0\_infrastructure\_bipartisan\_bill\_families, 2\_mask\_booster\_covid\_n95, etc. \\
C & 0\_right\_abortion\_must\_fight, 12\_registration\_register\_nationalvoter, etc. \\
D & 0\_trump\_covid\_relief\_need, 0\_asian\_american\_relief\_check \\
E & 1\_israel\_terrorist\_hamas\_palestinians, 1\_lgbtq\_pride\_month\_carl \\
F & 1\_tovah\_mar\_kippur\_yom, 1\_midnight\_book\_washington\_democracy \\
G & 1\_gaetz\_matt\_firebrand\_episode, 28\_firebrand\_episode\_matt\_gaetz \\
H & 1\_voter\_vote\_georgia\_democracy, 1\_officer\_capitol\_evans\_police \\
I & 3\_birthday\_friend\_daca\_wishing, 1\_abe\_japan\_minister\_prime \\
J & 36\_israel\_nuclear\_usa\_axis, 1\_israel\_isaac\_herzog\_lebanon\_breakthrough \\
\bottomrule
\end{tabularx}
\end{table}
\FloatBarrier

\section*{D. Moral Foundation Test Statistics by Longevity}

\begin{table}[H]
\caption{Test statistics comparing moral foundations across longevity levels}
\label{tab:mannwithney-mf-long-all}
\centering
\small
\begin{tabular}{l l c}
\toprule
\textbf{Longevity Comparison} & \textbf{Moral Foundation} & \textbf{p-value} \\
\midrule
High vs. Medium & Care & .0009 \\ & Fairness & .7817 \\ & Loyalty & .0072 \\ & Authority & .7023 \\ & Purity & .2059 \\
\midrule
Medium vs. Low  & Care & .0009 \\ & Fairness & .7817 \\ & Loyalty & .0072 \\ & Authority & .7023 \\ & Purity & .2059 \\
\midrule
High vs. Low    & Care & .0009 \\ & Fairness & .7817 \\ & Loyalty & .0072 \\ & Authority & .7023 \\ & Purity & .2059 \\
\midrule
High/Medium vs. Low & Care & .0009 \\ & Fairness & .7817 \\ & Loyalty & .0072 \\ & Authority & .7023 \\ & Purity & .2059 \\
\bottomrule
\end{tabular}
\end{table}
\FloatBarrier

\clearpage
\section*{E. Interactive Web App for Topic Exploration}

To complement the static tables and figures, we built a small interactive web
application that lets readers select a \emph{Congress member} and a \emph{topic}
to explore that member’s tweeting activity over time. The app displays:
(i) the member’s profile card and a monthly activity chart for the selected
topic, and (ii) the member’s ranking position on that topic relative to peers.
This interactive view helps connect the topic–frequency patterns discussed in
the paper with specific members and time windows.

\begin{figure}[H]
  \centering
  \includegraphics[width=\linewidth]{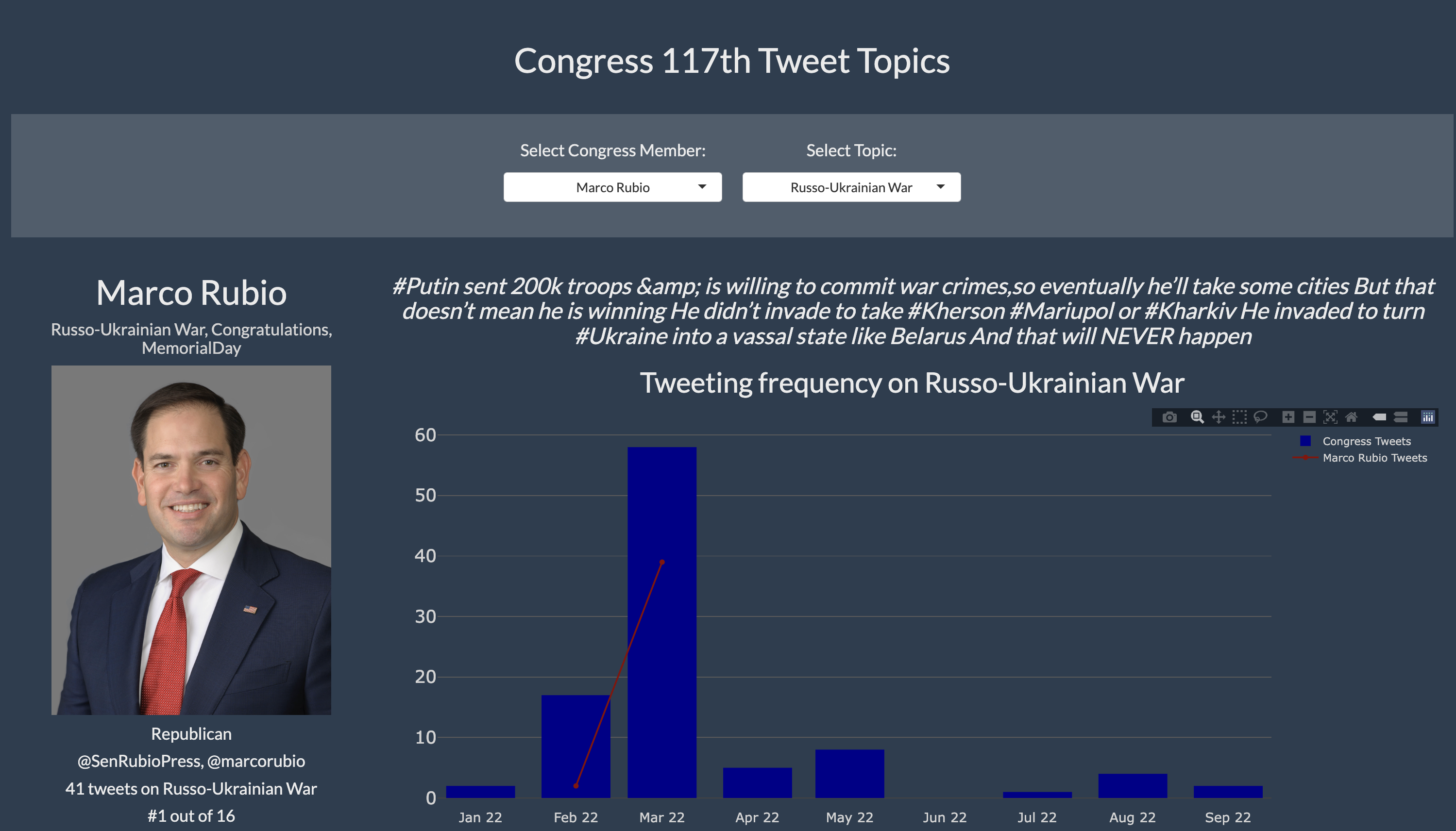}
  \caption{Interactive app: member/topic selection and monthly tweeting activity for the chosen topic.}
  \label{fig:webapp-activity}
\end{figure}

\begin{figure}[H]
  \centering
  \includegraphics[width=\linewidth]{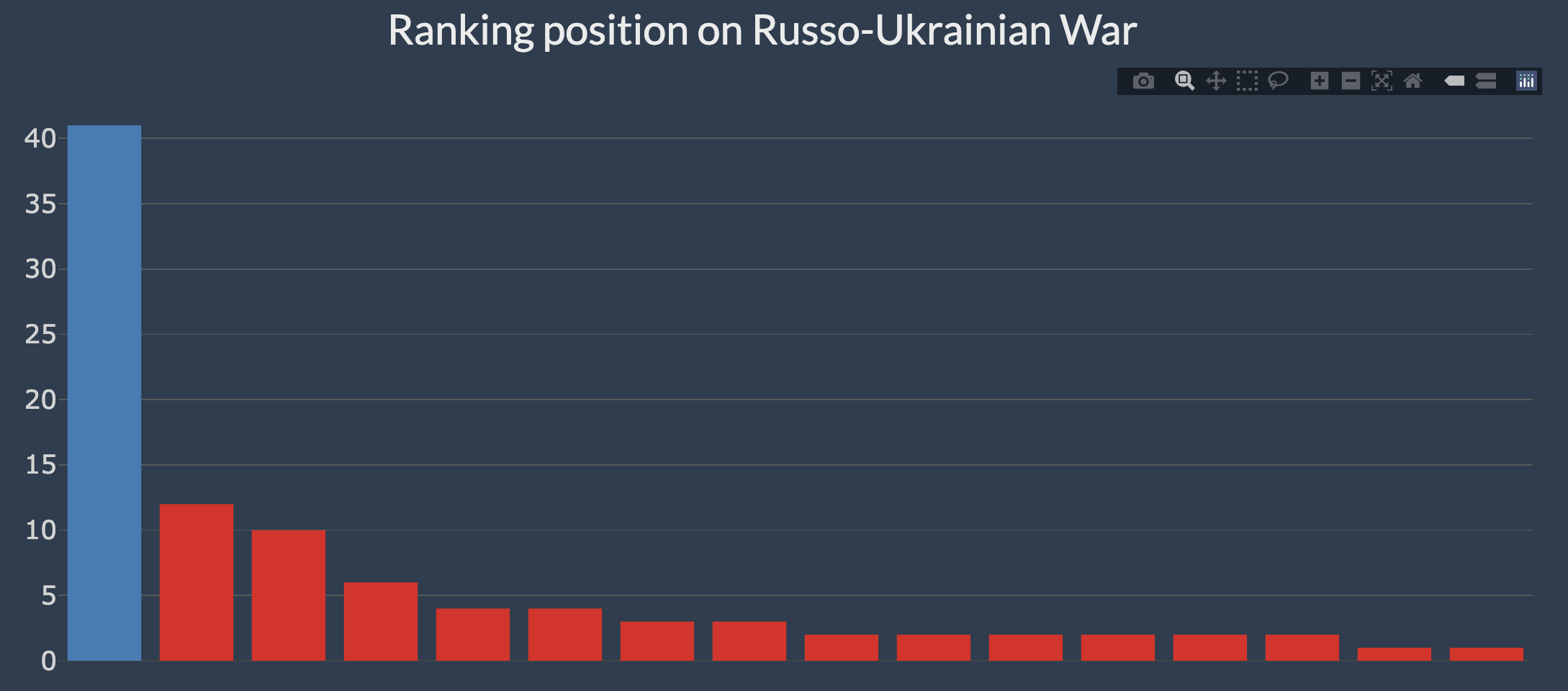}
  \caption{Interactive app: ranking position of the selected member on the chosen topic, relative to other members.}
  \label{fig:webapp-ranking}
\end{figure}

\FloatBarrier


\end{document}